\documentclass[aps,prl,twocolumn,groupedaddress,showpacs]{revtex4}

\usepackage{graphicx}% Include figure files in eps format.
\usepackage{amssymb}
\usepackage{amsmath}

\begin{document}

\title{An Experimental Investigation of the Scaling of Columnar Joints}

\author{Lucas Goehring}
\email[]{goehring@physics.utoronto.ca}
\affiliation{Department of Physics,
University of Toronto 60 St. George St., Toronto M5S 1A7, Ontario, Canada}
\author{Zhenquan Lin}
\email[]{linzhenquan@yahoo.com.cn} \affiliation{Dept. of Physics, Wenzhou
University, Wenzhou 325027, Zhejiang, China}
\author{Stephen W. Morris}
\email[]{smorris@physics.utoronto.ca} \affiliation{Department of Physics,
University of Toronto 60 St. George St., Toronto M5S 1A7, Ontario, Canada}

\date{\today}

\begin{abstract}
Columnar jointing is a fracture pattern common in igneous rocks in  which cracks self-organize into a roughly hexagonal arrangement, leaving behind an ordered colonnade.  We report  observations of columnar jointing in a laboratory analog system,  desiccated corn starch slurries.   Using measurements of moisture density, evaporation rates, and fracture advance rates as evidence, we suggest an advective-diffusive system is responsible for the rough scaling behavior of columnar joints.  This theory explains the order of magnitude difference in scales between jointing in lavas and in starches.   We investigated the scaling of average columnar cross-sectional areas due to the evaporation rate, the analog of the cooling rate of igneous columnar joints.  We measured column areas in experiments where the evaporation rate depended on lamp height and time, in experiments where the evaporation rate was  fixed using feedback methods, and in experiments where gelatin was  added to vary the rheology of the starch.  Our results suggest that  the column area at a particular depth is related to both the  current conditions, and hysteretically to the geometry of the pattern  at previous depths.  We argue that there exists a range of stable column scales allowed  for any particular evaporation rate.
\end{abstract} 


\maketitle

\section{Introduction}

Columnar jointing, the spontaneous fracture of rock into regular  prismatic columns, was first reported in the scientific literature in  the 17th century\cite{RBSRS1693}.  Travelers' reports of the Giant's  Causeway in Antrim, Northern Ireland, excited speculation as to what  mechanism could produce such a startling, seemingly unnatural display  of order and symmetry. It is now understood that columnar joints are  common in lava flows found worldwide, in a wide variety of materials  and deposition conditions.   An example outcrop is illustrated in  figure \ref{jointing} (a).  In lava, columns can vary in scale  between tiny, cm-sized  colonnades\cite{Patil1984}, to massive  pillars more than 2 m in diameter.  In spite of several centuries of  attention, the ordering mechanism is still, like many fracture problems, not fully understood. 

The qualitative mechanism of joint formation is easily described, but  questions of what sets the column scale and drives ordering are largely open.  The process typically begins with a pool of lava,  recently formed and immobile.  The lava can be assumed to have an  infinite horizontal extent, and cools from both upper and lower  surfaces.  Considering only one of the free surfaces, we find that a  solidification front moves into the melt as it cools.  Stresses  parallel to the front, caused by thermal contraction, lead to a  series of fractures perpendicular to the front.   The fractures  advance intermittently, with the fracture tips confined to a layer  only a few cm thick that trails closely behind the solidus.  Thus,  the pattern can be thought of as a dynamic 2D fracture network, that  leaves a columnar record of itself as it advances in the third  direction. As the cracks advance into the cooling rock, their pattern  self-organizes into a network that is deceptively close to, but not  quite, a regular hexagonal lattice \cite{Goehring2005}. Further  discussion of the geological process can be found in refs. \cite {Budkewitsch1994,Degraff1989}.

The pattern is difficult to study in the field, due to the  requirement of visualizing large landscapes in 3D.  However, in the  absence of strong observational constraints, a number of distinct  theories have been developed to explain columnar jointing.  In  general, these studies are interested in either scaling (refs. \cite {Reiter1987, Grossenbacher1995} for example) or ordering (eg. {\cite {Jagla2002, Jagla2004, Budkewitsch1994}), but rarely produce  predictions for both.  Fortunately, an analog columnar fracture  process can be observed in desiccated starch slurries, a fact that  was first systematically studied by M{\"u}ller\cite{Muller1998},  although it had evidently been noticed much earlier\cite{Huxley1881,  French1925}. In starches, the subject of this paper, columns are  usually between 1-10 mm in diameter (2-3 order of magnitude smaller than are typical in lava), as shown in figure \ref {jointing} (b).  This laboratory system offers an unusual opportunity  to do modern, controlled experiments that can test the various  theories, and perhaps shed light on this classic problem in  geomorphology \cite {Muller1998,Muller1998b,Muller2001,Toramaru2005,Goehring2005,Mizuguchi2005}.

\begin{figure}
\includegraphics[width=3.375in]{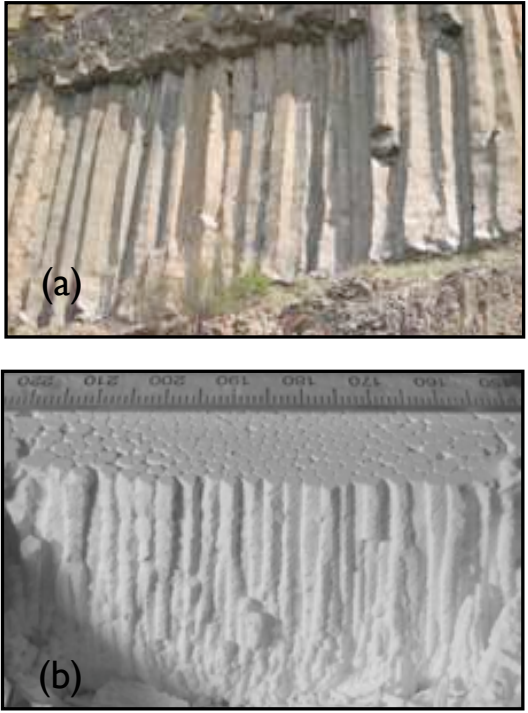}
\caption{\label{jointing} (Color online) Columnar jointing in (a) basalt of the Columbia Plateau near Banks Lake (95 cm average diameter), and (b) in desiccated corn starch.}
\end{figure}

The dynamic nature of the columnar crack network is one of the  features that has stimulated renewed interest in this old problem.   It has been suggested that there is a close relationship between a  number of different types of fracture patterns in which the fracture  network is able to evolve \cite{Goehring2005, Jagla2004,  Degraff1987}. Thin isotropic media undergoing brittle fracture due to  shrinkage typically display fracture networks which have  predominantly $90^\circ$ junctions and a mainly four sided polygonal  pattern of cracks\cite{Shorlin2000,Bohn2005}.  Evolved patterns, on  the other hand, often contain mostly $120^\circ$ junctions and  hexagonal units.  For example, ice-wedge polygons, which cover much  of the arctic terrain of the Earth, and also appear to cover  significant areas on Mars \cite{Mellon1997}, are seasonal thermal contraction cracks in permafrost.  The cracks open during the winter and heal during summer as water and/or detritus fills in the fissures \cite{Lachenbruch1962}.  The annual cycle allows the cracks to move slightly between years, and it has been observed that young ice-wedge networks contain more four sided polygons than older crack networks, which are roughly hexagonal  \cite{Sletten2003}.  Septarian concretions, roughly spherical bodies which have been fractured and filled in with another material, also can display similar hexagonal joint patterns \cite{Martinez1996}.

Columnar jointing itself also turns out to be a remarkably robust phenomenon.  Columnar jointing of some kind is found in many igneous rock types, such as basalt (including lunar basalt \cite{Jones}), rhyolite, andesite and dacite (see reference \cite{Degraff1987} for more detail).  It has been reported in other geological materials, such as sandstone and coal, and in man-made materials such as smelter slag, glass, chemically laced vitrified ice, and certain types of drying starches \cite{Seshadri1997,Menger2002,French1925,Muller1998,Degraff1987}.   Some igneous columnar formations can be very extensive.  An example is the Columbia River Basalt Group, which covers half of Washington state with columnar basalt hundreds of meters deep \cite{Alt1994}.  A good understanding of scaling could be used to extract quantitative information on cooling rate or heat fluxes from such igneous formations through simple geometric measurements.

Furthermore, columnar jointing is also interesting as a model problem that tests our understanding of fracture in general.  It is related to other interesting model problems such as thin film fracture (which should be the initiating pattern at the free surface), and directional drying of thin layers, which produces an analogous 2D dynamic ordering of cracks.  Both of these are subjects of intensive study \cite{Allain1995,Hull1999, Dufresne2003, Shorlin2000, Bohn2005, Bohn2005b}, and are important in many engineering applications of coatings and thin colloidal films. 

In this paper, we will report our recent experiments in desiccating  starch. Corn starch is just one of several types of starch which form  columns, but we have found it to be the most amenable to laboratory  experimentation.  We have developed several fully 3D visualization techniques to observe the fracture pattern, as well as methods to  systematically control the drying process \cite{Goehring2005}.  Using  these techniques, we focus in this paper on a number of tests of the  scaling and ordering mechanism of the starch columns.  We study  scaling within individual colonnades, and throughout series of  colonnades grown under a variety of drying situations, and attempt to  interpret the experimental results in terms of some general  theoretical considerations.  We also present observations of the  coarsening dynamics underlying the evolution of the observed column  scale, which suggests some insights into the ordering mechanism.

\section{Materials and Methods}

\begin{figure*}
{\includegraphics[width=7in]{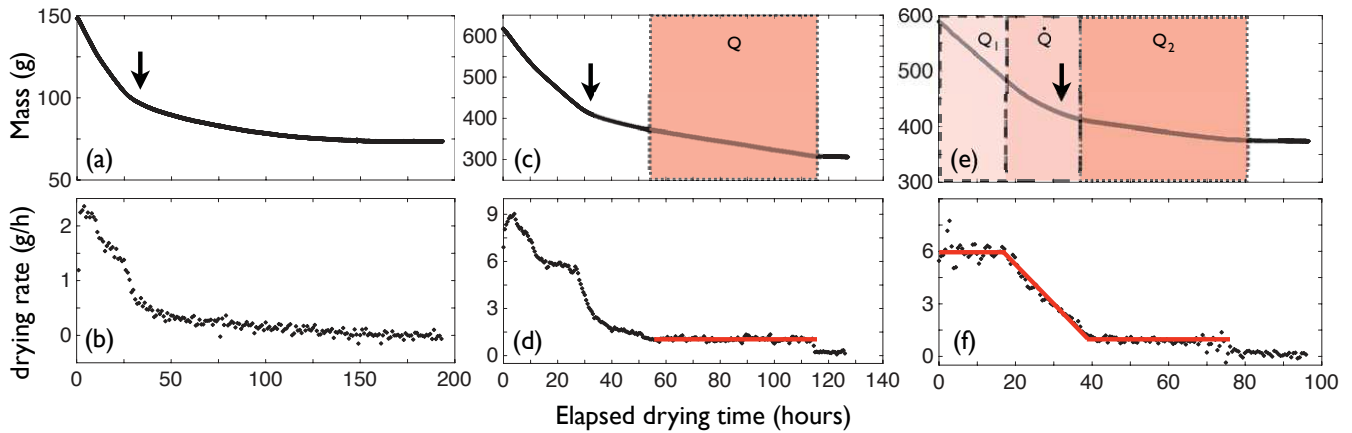}}
\caption{\label{drying} (Color online) The total mass of the drying  sample, {\it vs.} drying time.  (a) shows an uncontrolled  desiccation, with the evaporation rate (b).  (c) shows a feedback  controlled run in which the final drying rate in (d) is made  constant.  The three stage feedback controlled desiccation shown in  (e) and (f) was designed to initiate and develop the columnar  jointing in a controllable way.  Arrows point out the kink in the  curves that occurs when the sample ceases to dry homogeneously. In  (c)-(f) the dotted lines and boxes indicate controlled parameters.}
\end{figure*}

Experiments were performed in mixtures of corn starch and water.  Our methods are similar to those described by M\"{u}ller \cite{Muller1998} and build on our previous work \cite{Goehring2005}.  We desiccated slurries of equal weights of dry corn starch (Canada brand) and water, in round, flat-bottomed Pyrex dishes (90 mm inner radius), under 250 W halogen heat lamps.  Traces of bleach were added to sterilize the experiments.  Samples typically took 1-7 days to dry, and were 3-40 mm thick.  In some experiments, small amounts of gelatin were added to the initial slurry, in order to stiffen the mixture. The gelatin (Knox brand) was first dissolved in boiling water, mixed with cold water, and then added to the slurry, which was allowed to set before the experiment began. The initial weights of water and starch were kept equal in these experiments, but no bleach was added, as it tended to react with the gelatin. 

Fracture in drying corn starch occurs in two distinct stages\cite{Muller1998, Goehring2005, Toramaru2005, Bohn2005}.  Initially, a few first-generation (or primary) cracks, penetrating the full depth of the layer, break the starch layer into large pieces.  Subsequent, smaller columnar fractures are then observed to propagate slowly into the layer from the top drying surface.  The experiment is complete when the fractures reach the base of the slurry, and can be seen to appear in the starch on the underside of the Pyrex dish.

After each sample was dry, a set of columns, chosen well away from the edges of the dish, was removed.   Samples of the colonnades were analyzed in cross-section.   To avoid edge effects, we focused on the inner half of the sample's area, and did not include columns adjacent to first-generation cracks \cite{Goehring2005,Muller1998}.  Layers of starch were successively and destructively removed by sawing away the surface of this sub-sample.  Images were obtained of the surface after each cut, and cross-sectional areas were measured using Scion Image analysis software.  Typically, 100-200 polygons were measured for each layer, and, where shown, error bars represent the standard error of the measurements of the mean area. This destructive sampling technique produced useful column measurements at 2-3 mm depth intervals, but was too aggressive to use closer than $\sim$5 mm from the drying surface, where the structure is too delicate.  When higher resolution or near-surface observation was desired, we used micro-computed x-ray tomography (Micro-CT) to image starch colonnades after desiccation.  This technique provides full 3D volume filling images, as described in Ref. \cite{Goehring2005}.  Finally, in experiments where only the final cross-sectional area was wanted, we simply photographed and analyzed the base of the colonnade after removing it from its container.  When destructive sampling was used, this basal cross-section is also included as the terminal data point, to show the consistency of the methodology.

In order to make sense of the scale of the colonnade, it is essential to examine the drying process using controlled variations of the evaporation rate.  We studied the effect of several feedback  control schemes of increasing sophistication, shown in Fig.~\ref{drying}. In all cases where evaporation rates are given below, experiments were performed in 60 mm radius dishes. The mass of the slurry was measured once per minute by an electronic balance.    In the simplest experiments, the drying rate of the slurry was varied by manually adjusting the height of the heat lamps and the level of ventilation between experiments.  The relative heating intensity was measured with a photometer.  In these experiments, after an initial rapid homogeneous drying phase, we observed that the evaporation rate decreased smoothly with time, as shown in Fig.~\ref{drying}(a,b). Hereafter, we refer to these as ``uncontrolled" experiments.  To more directly control the evaporation rate throughout the experiment, we designed a computer-controlled feedback loop that allowed us to modulate the duty cycle of the heat lamps, and a small overhead fan. We used this technique to mimic the primary homogeneous stage of the uncontrolled evaporation curve as much as possible, using two types of feedback algorithm.  In one version, we controlled only the final evaporation rate, Q, as shown in Fig.~\ref{drying}(c,d).  For these experiments, the controlled evaporation begun slightly after columnar jointing was initiated.  In order to study the initiation of the pattern more carefully, we also implemented a three stage control algorithm,  as shown in Fig.~\ref{drying}(e,f).  This control scheme began with a period of constant, rapid evaporation, followed by a ramp down to a controlled final evaporation rate. The control algorithm has three free parameters,  the initial drying rate $Q_1$, the final drying rate $Q_2$, and the slope of the transition between them, $\dot{Q}$.  The initial water fraction for all runs was 0.50, by weight.  The $\dot{Q}$ ramp was started when the water fraction was reduced to 0.33.  Columnar jointing typically began late in the ramp phase.  In all feedback controlled experiments, the control loop terminated when the sample was fully dried, or when the duty cycle reached 100\%, when the algorithm was no longer able to maintain the desired evaporation rate.

\section{Results}

As can be seen in Fig.~\ref{drying}(a,b), we observed a multi-stage evaporation curve under uncontrolled conditions. This agrees with the results of Toramaru and Matsumoto \cite{Toramaru2005}, but contradicts M{\"u}ller, who had suggested that the desiccation curve was smooth \cite{Muller1998}.  We found that the change in the uncontrolled drying curve, marked by the arrow in Fig.~\ref{drying}(a), signals a transition between homogeneous and frontal drying.  Each has its own style of fracture; the former accompanies the primary cracks, while the latter accompanies columnar joint formation.  
In the initial stage of the desiccation, the slurry dries uniformly, and first-generation cracks open. The plumose structure\cite{Degraff1987} of these cracks shows that they initiate at the upper drying surface, and quickly propagate vertically through the entire sample. The resulting disordered fracture networks scale like those of ordinary thin films, and have been studied as examples thereof \cite{Bohn2005, Bohn2005b}.  In particular, the average spacing between first-generation cracks increases roughly linearly with sample depth, and the cracks typically meet at $90^\circ$ junctions.

When the whole sample reaches a water fraction of $0.30 \pm 0.01$, by weight, the slurry begins to dry from the top down.  This moisture content corresponds to the kink in the drying curve indicated by the arrows in figure \ref{drying}.  Columnar jointing is initiated from the drying surface at this time and further drying propagates a fracture front into the sample.  The front is a clearly delineated boundary between unfractured starch with a consistency of wet clay, and starch which has given way to columnar jointing. The two layers come apart naturally when a partially dried sample is removed from its dish, as all the fracture tips are confined to a thin interface, and the thickness of the columnar layer can be measured directly with a ruler.

Fig.~\ref{depthtime} shows the depth to which columnar jointing had progressed in the centre of the sample, in two experiments.  Each point represents columnar depth achieved in a separate partially completed desiccation.  Fig.~\ref{depthtime}(a) used uncontrolled evaporation with lamps 15 cm above the dishes, while Fig.~\ref{depthtime}(b) used a fully controlled evaporation with $Q_1$ = 4 g/h, $Q_2$ = 1 g/h, and $\dot{Q}$ = 0.5 g/h$^2$.  Both experiments had a fully dried depth of 33 mm.  In the uncontrolled run, the fracture front slows down as it enters the sample. We fit the position of the fracture front to a power law form with three free parameters, the amplitude,  exponent and the intercept on the time axis.  The uncontrolled experiment had an exponent of 0.60 $\pm$ 0.02, which implies that the slowing front penetrates slightly faster than normal diffusion.  On the other hand,  the corresponding fit  to the controlled experiment gives an exponent of 0.97 $\pm$ 0.06, indicating that controlling the evaporation rate fixes the average speed of the fracture front so that it progresses linearly into the sample.

\begin{figure}
\includegraphics[width=3.375in]{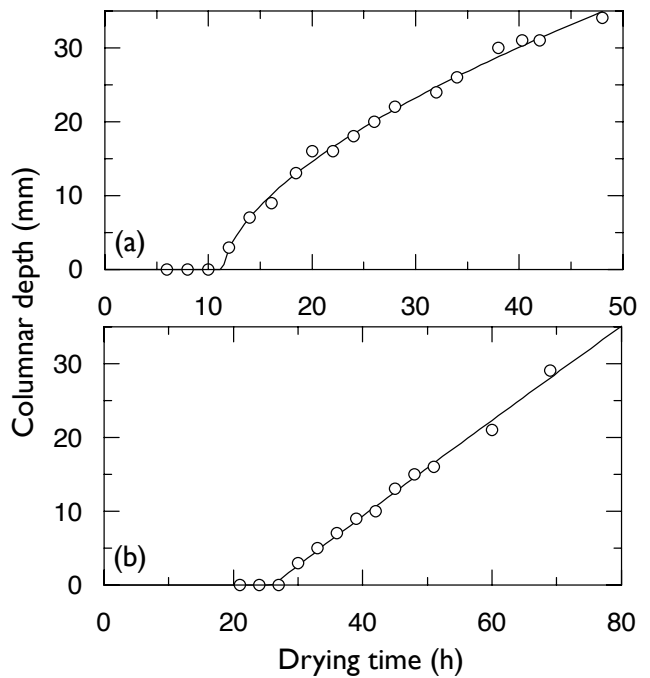}
\caption{\label{depthtime} Propagation of the fracture front into (a) uncontrolled, and (b) controlled starch desiccation experiments, with power law fits.  Errors ($\pm$1 mm) are approximately the size of the data points.}
\end{figure}

The columnar fracture front that we observed tracks the intrusion of a sharp moisture front.  By cutting a partially dried sample into thin layers, we measured the moisture content as a function of depth by subtracting the weights of small samples taken at various depths before and after baking them dry.  Fig.~\ref{water} shows the depth dependence of the moisture content of a starch sample 32 h into an uncontrolled desiccation.  A similar sharp drying front has also been seen in directional drying of 2D suspensions \cite{Dufresne2003}, and has been imaged in drying starch slurries using MRI \cite{Mizuguchi2005}.   Starch taken from below the fracture front closely matches the 0.30 moisture content which it achieved before the columnar phase of the desiccation.  Above the fracture front, there is a steep decline in moisture, starting at a sharp drying front.  The front is probably maintained by flow driven by capillary pressure; the ``wicking", of water towards the drying surface, rather than by solely by 
diffusion.  These effects are likely what lead to the non-diffusive power law behavior seen in the fracture advance rate.

To test that transport did in fact occur through the bulk, and was not facilitated by evaporation through the cracks, two experiments were set up using CoCl$_2$ and CuSO$_4$ indicators.  6 g of salt were dissolved in 100 g of water, mixed with 100 g of dry starch, and the slurries were dried.  The location of the salt deposition indicated where active evaporation was occurring.  In both cases, the indicators ended up at the upper drying surface of the desiccated sample, mostly within the top 2 mm. CoCl$_2$ is cobalt blue in its anhydrous state (when dry or warm), and red in its hydrated state (under cool, moist conditions).  In the case of CoCl$_2$, the entire sample of dried starch was homogeneously tinted light pink (likely due to residual moisture bound to the starch), while the surface was encrusted with a deep blue deposit.  In no case was salt left on the walls of the cracks, indicating that evaporation did not occur through the cracks.

These observations imply that {\it vapor} is not being transported up the cracks to the top surface.   This result agrees with that of Dufresne $\it{et \ al.}$\cite{Dufresne2003} who observed the drying of nanosphere/water slurries using CARS microscopy and concluded that that evaporation from the drying edge dominates over evaporation inside the cracks, where the humidity is easily saturated.  It remains possible, however, that cracks can have some enhancing effect on the {\it liquid} water transport \cite{Mizuguchi2005}. 

\begin{figure}
\includegraphics[width=3.375in]{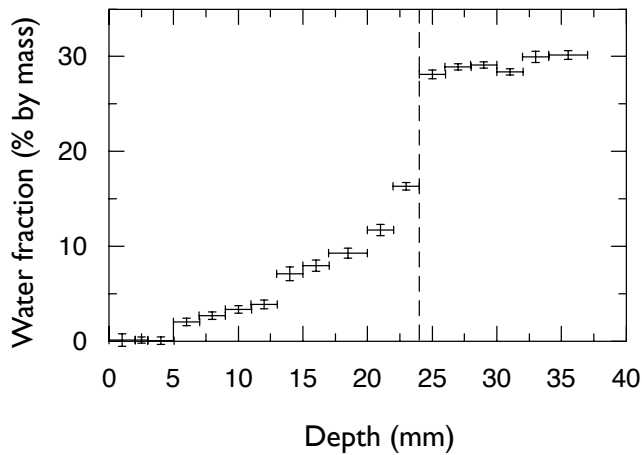}
\caption{\label{water} Fractional moisture content of a partially dried starch colonnade. The dashed line indicates the depth of columnar fracture. }
\end{figure}

We have previously reported our initial investigation of scaling behavior in starch colonnades.  Some of our results are summarized in figure \ref{depthscaling}, and more details can be found in ref. \cite{Goehring2005}.  These results allow us to estimate that errors in experimental repeatability between uncontrolled runs is, at most, approximately $\pm$8$\%$.  The statistical uncertainty in each measurement is smaller than this, approximately 4\%, and is about the size of the data points we present.  We found these results to be highly repeatable, and not dependent on the absolute depth of the sample.  For example, we performed two otherwise identical experiments on 7.5 and 5.5 cm thick slurries, and found that the position of a jump in columnar scale, as shown in Fig.~\ref{depthscaling} for the thicker experiment, occurred at the same depth in both cases.  In general, for uncontrolled drying, column scale increases, or coarsens, with depth \cite{Muller1998, Goehring2005}.  We showed that two types of coarsening were possible: smooth power-law-like increases, or sharp, almost discontinuous changes in column scale.  Using data gathered by 3D x-ray Micro-CT visualization, we observed a non-zero limiting column area near the drying surface.  

\begin{figure}
\includegraphics[width=3.375in]{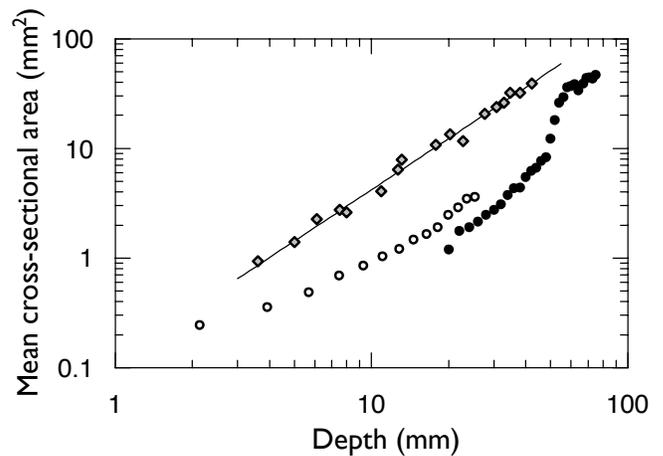}
\caption{\label{depthscaling} Average areas of corn starch columns under uncontrolled drying conditions, {\it vs.} depth.  The grey diamonds show the scales reached at the base of 17 identical slurries dried under the same, relatively slow conditions, but with different initial thicknesses. Micro-CT x-ray observations on a 25~mm deep, more quickly dried slurry (open circles) show a non-zero average column area near the drying surface. The filled circles show measurements from the destructive sampling of a thick, rapidly dried slurry that exhibits a sudden jump in average column area.}
\end{figure}

Building on our previous studies, we have observed identical dishes of starch slurries dried under heat lamps at various heights, but without feedback control.  Using a photometer, we measured the light intensity incident on the drying surface, which we assumed to be proportional to the incident heat flux.  Samples 6.5 mm and 13 mm deep were studied, and the mean cross-sectional area of the columns was measured at the base of the dried slurry. Fig.~\ref{lin1} shows the mean column scale as the lamp height was varied from 12-80 cm.  Faster drying generally gives rise to smaller columns.  For shallow samples, we find an approximately inverse relationship between column scale and lamp height, in agreement with Toramaru and Matsumoto\cite{Toramaru2005}, who observed, in similar experiments with potato starch, that the final cross-sectional area and the initial desiccation rate were inversely related. However, a strict inverse proportionality is only found for some of the data in the 6.5 mm thick samples.

\begin{figure}
\includegraphics[width=3.375in]{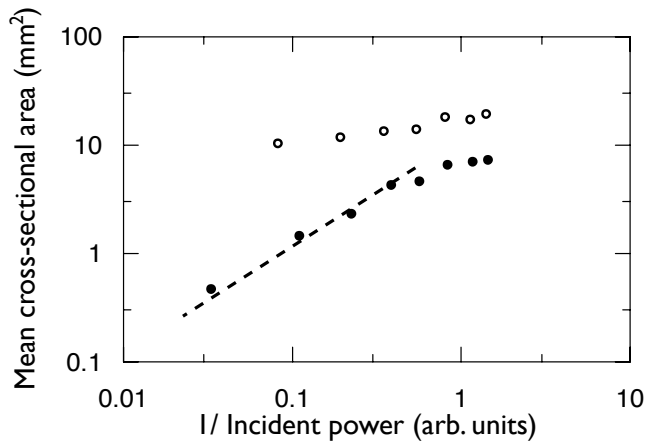}
\caption{\label{lin1} Average column areas at the base of the colonnade for 13 mm (open circles) and 6.5 mm (filled circles) deep samples.  The incident power records photometer readings as the heat lamp height was varied.  The line represents an inverse relationship between drying power and cross-sectional area.}
\end{figure}

We also investigated the effects of adding gelatin to the initial slurry.  This was done to place some qualitative experimental constraints on ball-and spring models of columnar joint formation.  Such ``discontinuum" fracture models \cite{Jagla2002, Kitsunezaki1999} are based on networks of linear springs with finite breaking strengths.    Fig.~\ref{lin2} shows that as gelatin is added, the average final column area increases, all else being equal.  This is what would be expected from discontinuum models since the added  gelatin presumably enhances the Young's modulus and fracture toughness of the networks.  More detailed measurements of the rheology of gelatin-starch mixtures, and more realistic modeling \cite{Tang2007} would be required to pursue this idea further. 

\begin{figure}
\includegraphics[width=3.375in]{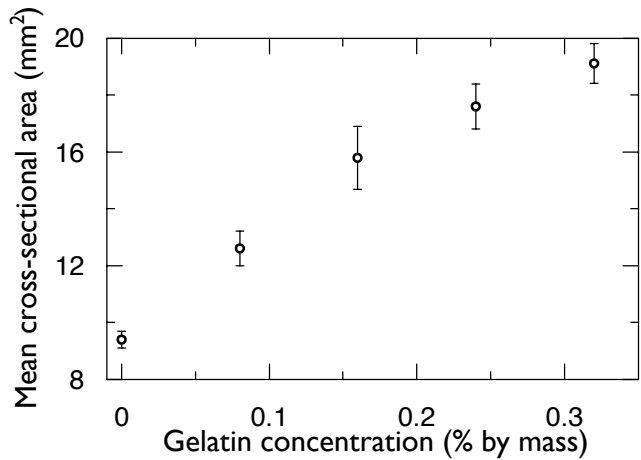}
\caption{\label{lin2} Average cross-sectional area at the base of five 17 mm deep samples, dried slowly with heat lamp 75 cm above.  Gelatin concentrations are given relative to the initial mass of the slurry.}
\end{figure}

The experiments just described show that even the simplest level of control has some reproducible effect on the final column scale.  The column scale was still coarsening in the interior of these samples, however.  This coarsening is linked to the slowing down of the fracture front during uncontrolled desiccation, as was shown in Fig.~\ref{depthtime}a.

To study the scaling behavior of corn starch columns under more steady-state conditions, we used automated feedback to fix the final desiccation rate.  In the single-control scenario, the overhead lamps were given an initial duty cycle of 0.5, and a feedback loop was switched on when the desiccation rate dropped to some desired value, $Q$.  The feedback loop then fixed the desiccation rate through the rest of the experiment, as shown in figure \ref{drying}(c-d).   In all cases, a constant final column scale was eventually chosen, and the coarsening that characterized the uncontrolled experiments was halted.  Figure \ref{single} shows the results of varying $Q$ from 0.9 to 1.5 g/h.  It also includes one experiment, where $Q$ = 0.7 g/h, but with an initial duty cycle of 0.25.  We found the results of these experiments to be rather surprising - all four samples dried under the same initial conditions, but with different final desiccation rates, were indistinguishable within errors.  The sample that was initiated differently, but subject to the same type of feedback during the controlled phase, selected a column area that was remarkably different. 

This puzzling observation suggested that perhaps long-term exposure to heat during the initial phase was slowly changing the physical properties of the starch slurries.  If this was true, the slower initiation of the 0.7 g/h sample could have resulted in a sample that had a different rheology than the others.  We were able to eliminate this possibility, however, by preparing six identical dishes of slurry, placing them all under one lamp, covering all but one with a Pyrex lid, and removing the lid from one additional sample every day.  The cross-sectional areas at the base of all six samples were identical within error.  Thus, it appears that the subsequent columnar fracture behavior of the samples is not affected by how long they are initially heated without drying.   We conjecture that all samples that have evolved up to the end of the uniform drying stage have identical water content and rheology.

\begin{figure}
\includegraphics[width=3.375in]{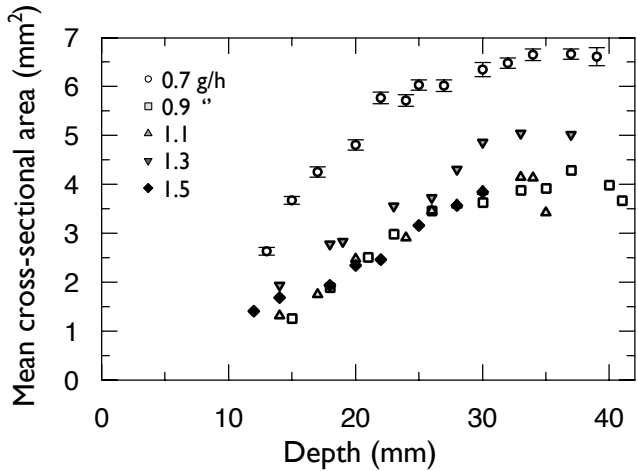}
\caption{\label{single} Average cross-sectional area of columns, measured through destructive sampling, for single-controlled experiments with a fixed final desiccation rate $Q$.  The initial duty cycle of the desiccating lamps on the 0.7 g/h sample is half that of the other samples. Representative errors are shown on the 0.7 g/h data.}
\end{figure}

Experiments using the most sophisticated three stage control algorithm confirm that the initiation and early development of a colonnade affects its scaling throughout the subsequent frontal drying regime.  In fully controlled experiments, the evaporation rate was fixed throughout the entire desiccation, as explained in figure \ref{drying}(e-f).  Fig.~\ref{double} shows the results of changing the control parameter $Q_1$,  leaving $Q_2$ and $\dot{Q}$ fixed.  Two data sets are shown, for which coarsening gradually comes to a halt as a final scale is selected. These experiments, which had the same final evaporation rate $Q_2$,  differ in final scale by over a factor of two.  The results show that, even given identical final evaporation rates $Q_2$, columnar desiccation can select average cross-sectional areas that vary considerably, depending on $Q_1$ and $\dot{Q}$.  We argue below that this reflects the intrinsically hysteretic nature of the fracture process.

\begin{figure}
\includegraphics[width=3.375in]{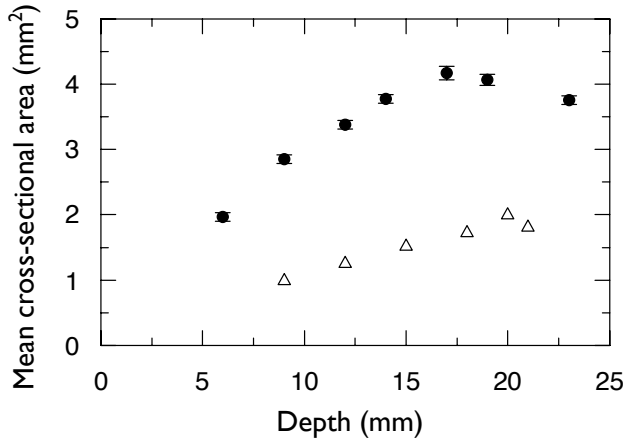}
\caption{\label{double} Average cross-sectional area of columns for two slurries under fully controlled desiccation rates, measured through destructive sampling.  The filled circles have $Q_1$ = 6 g/h, $Q_2$ = 1 g/h, and  $\dot{Q}$ = 0.5.  The open triangles are identical but use $Q_1$ = 9 g/h. Representative errors are shown on one data set.}
\end{figure}

By systematically varying $Q_2$ with the other drying parameters remaining fixed, we can explicitly study the response of average column area to a family of constant final evaporation rates.  Results for 600 g of initial slurry, $Q_1$ = 6 g/h and  $\dot{Q}$ = 0.5 are shown in Fig.~\ref{double2}.  The plotted cross-sectional areas are averages of 4-5 observations made from distinct cuts of destructive sampling, all near the bases of the colonnades.  The colonnades had a final depth of 33 mm, and average column areas remained constant in approximately the lower half of the colonnades.   Although there is a strong dependence of area on evaporation rate below $Q_2 =1$  g/h, there is little to no variation in area in the range $1 < Q_2 < 1.75$  g/h.  This shows that an established columnar pattern with a given scale can propagate stably over a range of evaporation rates.

\begin{figure}
\includegraphics[width=3.375in]{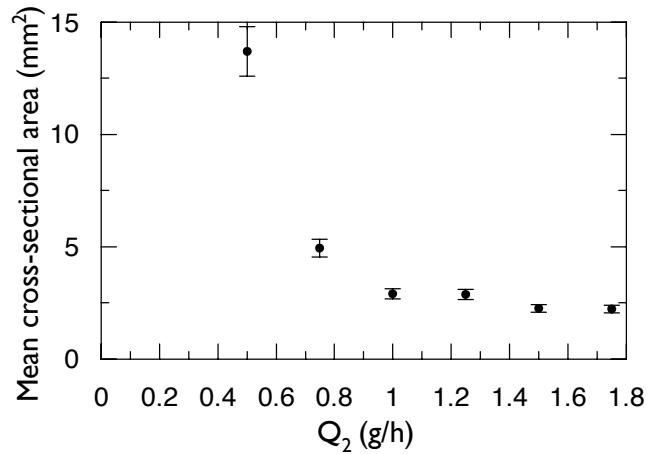}
\caption{\label{double2} Dependence of the average cross-sectional area of columns for experiments with $Q_1$ = 6 g/h and  $\dot{Q}$ = 0.5, with a varying final evaporation rate $Q_2$. }
\end{figure}

We have discussed three types of evolution that the average column area exhibits: average areas can smoothly increase with depth (i.e. coarsen), they can change scale catastrophically, or they can arrive at some constant,  ``selected" scale.  However, this list only describes the behavior of {\it average} column areas.  To study the underlying dynamics, we examined the evolution of single columns, using 3D x-ray Micro-CT tomograms for both uncontrolled and fully controlled desiccations \cite{epaps}.  

 Fig.~\ref{growth} shows the cross-sectional area for a few representative individual columns.  The mechanism of coarsening is the termination of joints as they intrude into the sample, resulting in the merger of two or three columns \cite{Goehring2005}.   When columns merge, which occurs at the discontinuities in Fig.~\ref{growth}, we continued to track the offspring column.  Figure \ref{growth} (a-b) shows that after a merger event, the column rapidly shrinks back to average size, giving up area to its neighbors.  A column that is  spawned from a vertex also influences its neighbors, growing at their expense.  In most cases, we found that such a new column merges with one of its neighbors relatively quickly, although occasionally one  survives and grows to average size.  An example is shown in Fig.~\ref{growth}(c).  The dashed arrow indicates the creation of a new column at a vertex of the column whose area is plotted. When this happens, the observed column shrinks as its new neighbor grows, until the two columns merge at the solid arrow.  In general, changes in an individual  column's area appear to involve purely local processes, and affect only the areas of its immediate neighbors.

\begin{figure}
\includegraphics[width=3.375in]{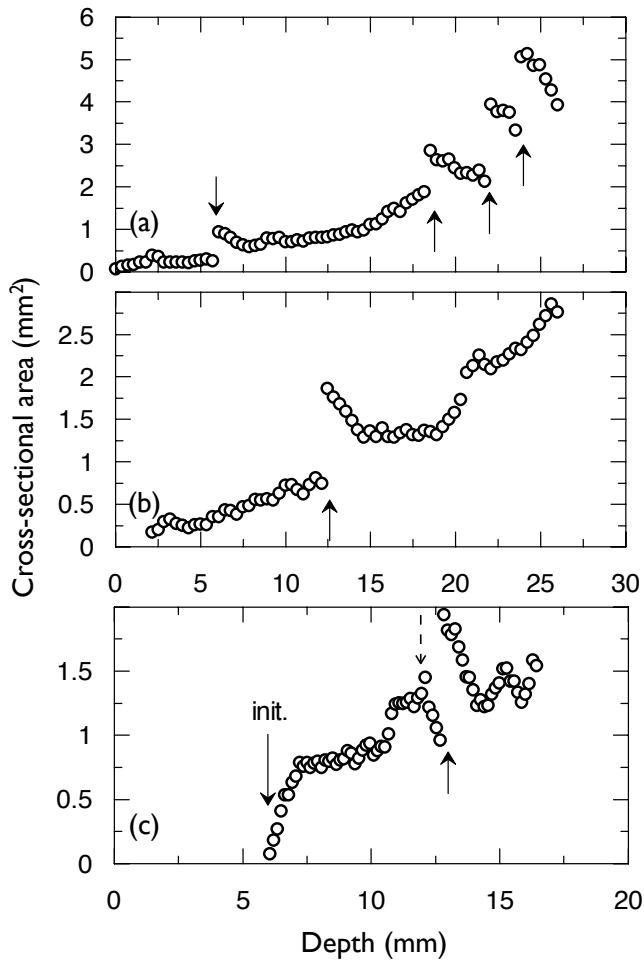}
\caption{\label{growth} Area evolution for selected individual columns in uncontrolled coarsening (a-b) and controlled (c) experiments.  Column merger events are marked by a dark arrow pointing to discontinuities in area. In (c), a dashed arrow points to the creation of a new column, which grew from of an adjacent vertex.}
\end{figure}

\section{Discussion}

The range of experiments discussed above present a rather complex picture and require careful interpretation. If we restrict ourselves to the question of how the average column area that is finally selected scales with final desiccation rate, we find that no single, simple answer emerges from the data.  
Taken together, however, the experiments make it possible to draw some important general conclusions about the mechanisms of scale selection in columnar joints in both starch and basalt. 

\subsection{Advection-diffusion}

The dynamics of columnar jointing in starch can only be be understood in conjunction with the dynamics of the moisture content within the sample.  Given the observations shown in Figs. \ref{drying}-\ref{water}, we suggest that a nonlinear diffusion equation such as
\begin{equation} \label{nonlinear}
\partial_t c = \nabla \cdot [D(c) \nabla c],
\end{equation}
where the hydraulic diffusivity $D$ depends on the water density $c$, is appropriate to model the water transport in the bulk of the starch.  We can treat transport by the cracks, if any, as part of the boundary conditions on this equation. This suggestion is in accord with the conclusions of Mizuguchi $\it{et \ al.}$, who performed MRI on desiccating samples \cite{Mizuguchi2005}.  The porous medium equation that can be obtained by from Eqn. \ref{nonlinear} by setting $D(c) \propto c$ is known to give rise to propagating self-similar solutions with parabolic shape \cite{Barenblatt1996}.  Analogously, the shape of the sharp front shown in Fig.~\ref{water} could be explained by the presence of a minimum in $D(c)$ at a moisture content of 0.3, by mass (a water density of $c_0$ = 0.33 g/ml).  Such a sharp front is expected when capillary forces become the dominant driving mechanism for water transport, and it indicates a shift from flow-limited to evaporation-limited dynamics \cite{Dufresne2003}.  Dufresne {\it et al.} \cite{Dufresne2003} showed that the cross-over length of this type of front scales inversely with the evaporation rate, analogous to our parameter $Q_2$. This length is characterized by the width of the dropoff in moisture at the desiccation front, and cannot be more than a few mm in our experiments. Such fronts are typical in drying slurries \cite{Dufresne2003, Salamanca2001, Allain1995, Pauchard2003}. 

We now consider a starch sample in which the crack front and the drying front both move at constant speed $v$ in the $z$ direction, as in the fully controlled experiments with constant evaporation rate. Moving to the co-moving frame introduces an advective term into Eqn.~\ref{nonlinear}, 
\begin{equation} \label{advection}
\partial_t c = v \partial_z c + \nabla \cdot [D(c) \nabla c].
\end{equation}

Around the minimum in $D(c)$ near the drying front, we can make the Taylor expansion
\begin{equation} \label{taylor}
D(c) = D_0+ D_2 (c - c_0)^2 + ...\ .
\end{equation}
The position of the crack tips follows closely behind the desiccation front, so that they will be subject to an effective diffusivity close to $D_0$, but the diffusivity will increase rapidly away from the position of the crack front.  Under these conditions, a constant moisture content boundary condition leading the front, and a constant flux boundary condition trailing the front are appropriate.  These will quickly lead to a time-independent version of the nonlinear advection diffusion equation, which assumes the nondimensional form,
\begin{equation} \label{nondim}
 \nabla ^2 c + {\rm Pe ^\prime}~\partial_{z^\prime} c = {\rm nonlinear~terms},
\end{equation}
where $z^\prime = z/L$, and the hydraulic P\'{e}clet number, ${\rm Pe^\prime} = vL/D_0$ measures the ratio of advective to diffusive effects.  We take the characteristic length scale $L$ to be the effective radius of a column.  

The hydraulic P\'{e}clet number offers some interesting insights into the relative scale of corn starch columns and lava columns.  For a typical controlled experiment with a fixed evaporation rate of 1 g/h, we found a fracture velocity of $v = 0.18\pm 0.01\ \mu$m/s, as shown by the slope of the data in Fig.~\ref{drying}b.  The observed average cross-sectional area for these conditions was 2.9$\pm$0.2 mm$^2$ using the data from Fig. \ref{double2}.  However, as shown in Fig. \ref{double}, changes in the initiation of the drying process ({\it i.e.} in $Q_1$) can cause this area to be vary between at least 1.5-4 mm$^2$.  We therefore choose a value of  $L$ in the range 0.7-1.1 mm.   

The diffusion constant $D_0$ is the most difficult property to measure, especially since we do not know what the effect of the cracks is on the water transport.  If we make the plausible assumption that it is small (as is certainly true for the vapor transport \cite{Dufresne2003}), then we can estimate $D_0$  by fitting a square root function to an uncontrolled desiccation curve. Assuming that the position of the fracture front, $\chi$, tracks a particular concentration $c_f$, and that the minima in diffusivity near the desiccation front dominates the dynamics, we can estimate the position at any time using a solution of the diffusion equation,
\begin{equation} \label{iso}
\chi(t) = \sqrt{4D_0 t} \  {\rm erf}^{-1}(c_f/c_0).
\end{equation}
which describes diffusion from a half-space with a dry dirichlet boundary condition.  While not strictly valid in this case, it is nonetheless a reasonable first estimate of $D_0$.  Given the uncertainties in $c_f$ from the moisture content curve, we estimate $D_0$ = (1-5)$\times10^{-9}$ m$^2$/s.  This is somewhat lower than the estimate of the diffusivity by M\"{u}ller, who obtained (0.7-2.7)$\times10^{-8}$ m$^2$/s \cite{Muller1998}.  However, M\"{u}ller's  value was based on measurements of mass changes during only the first five hours of evaporation, which probably leads to an overestimate.  In fact, it is likely that the best value of $D_0$ is near the low end of the stated range.  Combining these observations, we estimate that columnar jointing in corn starch occurs at a value of  ${\rm Pe^\prime}  \sim$ 0.2, within a factor of 2.

The advection-diffusion model requires only slight modification to apply to the columnar jointing of lavas.  The thermal formulation of eqn. \ref{nondim} is simply
\begin{equation} \label{nondim2}
 \nabla ^2 T + {\rm Pe}~\partial_{z^\prime} T = 0,
\end{equation}
where $T$ is the temperature and ${\rm Pe} = v L/\kappa$.  The thermal diffusivity $\kappa$ of basalts does not vary by more than a factor of two for temperatures $0 \leq T \leq 1500^\circ$C \cite{Mostafa2004,Murase1973}, so the nonlinear terms in Eqn.~\ref{nondim} have been neglected. 

In the case of lava, it is known that heat transport by the cracks makes an important, indeed dominant, contribution to the total heat transport. The crack surfaces are treated as boundary conditions on Eqn.~\ref{nondim2}.    If the surface of the crack is maintained at a constant temperature $T_c$, then the crack heat flux is given in dimensionless terms by Newton's law of cooling,
\begin{equation} \label{biot}
\partial_{y^\prime} T  = {\rm Bi} (T-T_c), 
\end{equation}
where the Biot number ${\rm Bi} = h L/\lambda$.  Here, $\lambda$ is the thermal conductivity and $h$ is a phenomenological heat transfer coefficient.  The coordinate $y^\prime$ is perpendicular to the crack surface. It is believed that cooling in fractured lava is greatly enhanced by the boiling and reflux of groundwater in the cracks \cite{Hardee1980, Budkewitsch1994}.   The high efficiency of this convective heat transfer process compared to that of thermal conduction in lava suggests that the Biot number is large.  It is important to note  that, although the overall cooling is driven by the boiling of water, the fracture front in lava is located in a region at much higher temperature, near 900$^\circ$C \cite{Peck1968}.  The fracture tips are therefore likely to be in a region where ordinary thermal conductivity dominates, and up to several meters distant from the cooler convective zone.  
 
If, as we previously assumed, the cracks in the starch do not significantly enhance the water transport, then the corresponding ``hydraulic Biot number" is effectively zero.

We can now estimate the P\'{e}clet number ${\rm Pe}$ relevant to the formation of columnar joints in lava. 25 years of observations of the gradual solidification of the Kilauea Iki lava lake, formed in 1959, show that an Eqn.~\ref{nondim2} accurately describes the isotherms in boreholes and the position of the solidification front \cite{Hardee1980}.  These observations yield measured values $\kappa = 5.0 \times 10^{-7}$ m$^2$/s, and $v = 6.7 \times 10^{-8}$ m/s.  Observations of columnar jointing near the Kilauea caldera of Mauna Loa (although not on Kilauea Iki) suggest that columnar jointing at this site is typically rather large, with diameters in the range 1.5-3 m  \cite{Peck1968}.  From these data, we find ${\rm Pe} \sim $ 0.15, in remarkable agreement with the value of ${\rm Pe}^\prime$ we found for the starch experiments.  However, both values should only be considered accurate to within a factor of 2.

In summary, we find that the shrinkage fronts in both cooling lava and desiccating starch are similar, in dimensionless terms.  Both systems are controlled by advection-diffusion processes with similar P\'{e}clet numbers.  They differ, however, in several secondary aspects.  The transport in the bulk of the starch is probably by strongly nonlinear diffusion, while the heat transport in the bulk of the lava is linear.  The role of the cracks in transport likely also differs.  In the starch, it appears most likely that the cracks do not significantly enhance water transport. In the case of lava, groundwater boiling and thermal conduction combine to give a rather complicated boundary condition on the temperature and its gradient at the cracks.  In either case, however, the result of the transport is the propagation of a sharp shrinkage front.
The order of magnitude differences between the scales of joints in lavas ($\sim$1m columns) and starches ($\sim$1 mm columns) must be directly related to the dynamic similarity of the shrinkage fronts in the two cases.  By imposing a constant evaporation rate, we can experimentally probe the detailed relationship of the fracture scale to the speed of the shrinkage front.

\subsection{Scale selection, memory, and hysteresis}

It has often been implicitly assumed that there is a simple one-to-one relationship between the scale of columnar joints and some externally imposed driving force.  For joints in lava, it is generally agreed that there is some dependence of the column cross-sectional area on the cooling rate \cite{Degraff1989,Long1986,Budkewitsch1994,Reiter1987,Grossenbacher1995}.  The primary geological evidence for this relationship comes from studies of the crystal texture of the rock in columnar layers separated by regions of {\it entablature} (smaller scale, irregular jointing).  The crystal texture of the columnar layers is coarser, implying that it cooled more slowly than the entablature \cite{Degraff1989,Long1986}.  Further evidence exists in a rough correlation linking the average column area to the widths of fracture features known as {\it striae}, which are believed to be a proxy for cooling rate in basalt \cite{Ryan1978,Long1986,Degraff1993}.  In starch experiments \cite{Muller1998,Goehring2005,Toramaru2005}, it is also observed that faster evaporation rates generally give smaller columns.  While suggestive, these qualitative results are insufficient to supply any strong constraint, much less to imply a one-to-one relationship, between cooling or evaporation rates and average column area.  Nonetheless, the assumption that there exists such a relationship is implicit in much of the modeling of columnar jointing phenomena \cite{Reiter1987,Grossenbacher1995, Toramaru2005}.  

This assumption is not universally accepted.  M\"{u}ller, although using the one-to-one assumption in his analysis, expresses concern over the lack of experimental proof for this principle \cite{Muller1998}.  Saliba and Jagla \cite{Saliba2003} have produced a finite element calculation of a fracture front intruding into a cooling solid, and have made the alternate prediction that the fracture scale is independent of the driving temperature gradient --- except possibly during initiation.  We believe that our experiments are the first to examine the experimental relationship of columnar joint scaling to fixed gradients and their associated fixed fracture advance rates.

Our controlled experiments show that when the evaporation rate is explicitly fixed, there are some regimes for which there is no strong dependence of the selected average cross-sectional area on the final evaporation rate.  In other regimes, there is such a dependence.  Furthermore, there are ranges of column scales that are attainable via different approaches to the same final evaporation rate.  And finally, there is the observation of smoothly changing evaporation rates leading to sudden jumps in column scale \cite{Goehring2005}.

The experiments of Toramaru and Matsumoto suggested a simple inverse relationship between average area and the evaporation rate \cite{Toramaru2005}.  This result is particularly promising since we expect the width of the desiccation front to vary as 1/$Q_2$.  It is well known that uniformly dried, homogeneous thin films have a fracture spacing that is proportional to the layer thickness \cite{Groisman1994, Kitsunezaki1999, Shorlin2000, Bohn2005b}.  However, even though some of our experiments with varying lamp height yield similar results, they cannot be fit as easily to a purely inverse relationship.  As we showed in \cite{Goehring2005}, the average areas in these experiments would probably continue to evolve if the samples were thicker.  

All these observations can be made consistent with each other if we assume that the columnar joint pattern exhibits  a form of dynamic  stability, leading to hysteresis.  If the pattern has a memory, there may exist a range of column scales that are stable for any given evaporation rate.  The scale observed then depends on the evolutionary history of the pattern.

By suggesting that the pattern of columnar joints has a memory, we are claiming nothing more than that the fracture network at any given depth is, in general, similar to the fracture network at depths near to it.  This is suggested by the way that column scale changes.  For a small increase in average column area, a number of discrete, local column mergers occur, after which the column network slowly relaxes by propagating the change through local interactions of columns with their neighbors. 
Small, rapid changes over the whole pattern are not possible, or at least highly constrained by geometry.  Thus, a given pattern scale may be stable over some range of conditions.  Large, sudden changes of average column scale may be possible if a stability threshold is passed where, on average, it is more favorable to have all columns more than twice as large as they currently are.  Under these conditions, a large scale change can be enacted by a catastrophic event consisting of many column mergers, as we saw with the jump in columnar scale in Fig.~\ref{depthscaling}.  This has the appearance of a global instability of the whole pattern.

The experiments that we report in Fig.~\ref{double} imply that a range of column scales could be selected under identical final evaporation rates.  Alternatively, the data presented in Figs.~\ref{single} and \ref{double2} show that the same columnar scale can be chosen over a range of final drying rates, given the same drying initiation.  These results suggest a boundary between column scales that are stable, and those that coarsen, for a given evaporation rate.  Different drying histories could lead to the selection of different scales from within this stable range.  

A superficially appealing argument for a uniquely selected column scale under given external conditions has often been sought in energy considerations.  It has been asserted that the hexagonal geometry of columnar joints arises because such a form optimally releases the strain energy per unit area, or otherwise optimally reduces the free energy of the system \cite{Mallet1875, Jagla2002}.  In this view, the evolution of the columns tends toward a perfectly hexagonal lattice, or at least to a static pattern corresponding to a local free energy minimum with frozen-in disorder \cite{Jagla2002, Budkewitsch1994}. Our results suggest a more dynamical interpretation, in which a nonunique scale may be selected within a stable band, depending on system history. Such a pattern need not be perfectly ordered \cite{Goehring2005} or static \cite{epaps}. 

A quantitative approach to the energetics of fracture patterns is provided by Griffith fracture theory \cite{Griffith1921}. In Griffith's theory, fractures can only advance if the rate of increase in surface energy of the opening fracture is less than or equal to the rate of decrease in strain energy.  The difference, if any, between these rates represents dissipation; fracture is, in general, a thermodynamically irreversible process.   A simplifying assumption often made is to assume quasi-static, or steady, slow fracture advance, in which case the dissipation is presumed to be small and the equality in the Griffith's criterion holds. This is a rather restrictive and unrealistic assumption, however.  In fact, both the rate of strain energy generation and dissipation are proportional to the velocity of the fracture front.  Griffith theory can therefore only be used to set an upper bound on the fracture density, $\rho$
expressed as the crack area per unit volume, given by 
\begin{equation} \label{griffith}
\rho \leq \rho_{max}= \frac{\Delta \epsilon}{\gamma_s},
\end{equation}
where $\Delta \epsilon$ is the difference in the strain energy density between wet and dry samples, and $\gamma_s$ is the energy required to open a unit area of crack surface. Thus, in the presence of dissipation,  energy considerations alone are insufficient to require a one-to-one relationship between fracture advance rates and column scale.

Irwin's modifications to Griffith theory \cite{Irwin1957} are somewhat more useful to the problem of columnar jointing.  Irwin introduced the concept of fracture toughness, and showed how the stress intensification in the vicinity of a crack implies that it is much easier to lengthen an existing crack than it is to nucleate a new one.  Although a quantitative calculation of the stress intensity factor would be difficult for a geometry as complex as that of columnar joints, the existence of this factor can supply the effect of memory to the columnar structure.  For a wide range of stresses, the downwards extension of existing cracks will be favored over the nucleation of new ones, or the termination of old ones.  Thus, if a network of cracks with a given average area is advancing under changing conditions, their propagation may be stable, even if the changed conditions would not initiate a pattern of that scale.  
% wording changes
The discontinuous jumps we have observed could indicate the crossing of a stability boundary between regions of relatively unchanging crack propagation.

\subsection{2D columnar analogs}

Since no comprehensive theory exists for the 3D case of columnar joints, it is useful to consider better-studied 2D analogs. In these cases, propagating arrays of periodic, parallel cracks are observed in behind shrinkage fronts moving through thin samples.  Both thermal and desiccation experiments have been performed.  Crack arrays form when a thin layer of wet slurry is dried from one edge, both in layers with one free surface \cite{Shorlin2000,Pauchard2003}, and in suspensions confined between two plates \cite{Allain1995, Hull1999, Dufresne2003}.  The thermal experiments typically consist of dipping a heated strip of glass, or other material, into a cooler water bath at a fixed rate \cite{Yuse1993, Ronsin1997}.  These 2D fracture processes have also been studied theoretically in some detail \cite{Jagla2002, Boeck1999, Komatsu1997}.  

Boeck $\it{et \ al.}$ \cite{Boeck1999} have proposed a theory for the case of a thermally-quenched strip.  Beginning with Eqns.~\ref{nondim2} and \ref{biot} for the planar temperature field, and the appropriate equations for the temperature-dependent stress field, they assume quasi-static fracture advance with speed $v$ and crack spacing 2$L$.  Under these assumptions, they find a one-dimensional manifold of solutions linking $L$ and $v$.  Thus, for a given $v$, there {\it may} exist a unique spacing $L$ which is stationary in the co-moving frame, or there may not.  In general, what is actually observed depends on stability considerations.  Some stationary solutions will be unstable to small perturbations, leading either to new stable solutions with different $v$ and $L$, or possibly to no stable solution.  For a restricted set of perturbations, they analyzed a period doubling instability where every other crack falls behind the front and presumably stops. They showed that there exists a critical P\'{e}clet number ${\rm Pe}_c$ such that for  ${\rm Pe} > {\rm Pe}_c$, {\it any} value of the crack spacing is stable \cite{Boeck1999}.  ${\rm Pe}_c$ depends only very weakly on Bi, varying only by a factor of two when Bi is varied over four orders of magnitude.  While these calculations cannot be directly applied to the 3D case of columnar joints, they do suggest that all columns larger than some critical size might be stable for given conditions, and hence that similar stability criteria might drive the observed coarsening towards larger column scales.  The period doubling instability that Boeck $\it{et \ al.}$ considered bears some resemblance to the sudden changes in column scale we observed in rapidly dried starch \cite{Goehring2005}.

In another calculation, Jagla  \cite{Jagla2002} considered the mechanism by which crack arrays in  2D regularize their spacing under the action of the stress distribution.  A pair of cracks whose spacing is wider than neighboring pairs develops an asymmetric stress concentration near the crack tips.  When these cracks advance, they do so in a direction that reduces the maximum local stress, and thus curve inwards slightly, in accordance with the Cotterell-Rice principle \cite{Cotterell1980}.  This tends to produce equidistant cracks and hence a regular spacing.  The local equalization of column areas that we have observed in individual columns is presumably due to the action of analogous stress distributions, although their 3D morphology is much more complicated.

2D experiments also suggest that a range of crack spacing can be stable for the same external conditions.  In 2D, crack spacing generally increases linearly with sample thickness \cite{Shorlin2000,Pauchard2003}. Ronsin and Perrin \cite{Ronsin1997} observed a window of allowed crack spacings for a given dipping velocity in experiments with glass .  Similarly, Shorlin $\it{et\ al.}$ \cite{Shorlin2000} performed experiments on directionally dried slurries on substrates that incorporated a discontinuous step.  In certain cases, the original pattern of parallel cracks continued with the original spacing, even over step which doubled the layer thickness. In other cases, the cracks responded by doubling their spacing.

Thus, a variety of experiments and theory in 2D make it plausible that a hysteretic, non-unique column scale exists for columnar joints, within some band of scales limited by stability boundaries.  The task then is to locate those boundaries and understand the stability properties of the evolving fracture networks that lead to columnar structures.

\section{Conclusion}

We have studied the scaling behavior of columnar joints using three-dimensional observations of desiccated starch colonnades.  Our results are primarily experimental; we measured the evolution of the  average column area and the size of individual columns under a range of controlled and uncontrolled conditions. We found a relatively complicated relationship between the desiccation rates and the  average column area.  Individual columns rapidly evolved toward the average area, even while coarsening and merging dynamics continued.  We argued that the regularization of column areas and the stress intensity factor in the vicinity of existing cracks produces a kind of geometric memory, which allows scale selection to be history dependent.  We demonstrated this most clearly in the fully controlled experiments.  In some cases there was a selected scale that was stable over a range of final evaporation rates.  In other cases, where the final evaporation rate was fixed, changes in the initiation influenced the final scale that was selected.  These results imply that there is a window of column scales that are allowed under any given externally applied evaporation rate.

We have sketched a theory of the advective concentration dynamics in which the water moves by nonlinear diffusion.  By estimating the hydraulic P\'{e}clet number for this process, and comparing it to the thermal P\'{e}clet number for columnar jointing in Hawaiian lava lakes, we have shown a dynamic similarity that may explain the relative scale between columnar jointing in starches and in lavas.  To complete this theoretical framework, a theory of the rheology and fracture properties of the starch needs to be formulated, and compared to the corresponding theory for lava.

We argued that the scale of columnar joints is probably not fully deducible from the energy balance between the elastic energy released and the energy consumed to create new fracture surfaces.  Rather, the column scale and its dynamical evolution will require a more complete understanding of the stability and instability of propagating fracture networks with their associated strain fields and their diffusing moisture or temperature fields.  Such a theory, along with further controlled experiments on starch desiccation, as well as new field observations of lava columns, should enable some progress toward unravelling the puzzle of these strangely shaped rocks.

\section{Acknowledgments}
We wish to thank Mark Henkelman, and the Mouse Imaging Centre (MICe) for providing Micro-CT facilities.  This research was supported by the Natural Science and Engineering Research Council of Canada.   Support for Zhenquan Lin was provided under the National Natural Science Foundation of China under Grant No. 10275048 and the Zhejiang Provincial Natural Science Foundation of China under Grant No. 102067.  

\end{document}